\documentstyle[pra,aps]{revtex}
\begin{document}
\def\question#1{{{\marginpar{\tiny \sc #1}}}}
\draft
\title{Coherence of neutrino flavor mixing in quantum field theory}
\author{Christian Y. Cardall}
\address{ Department of Physics \& Astronomy, 
	State University of New York at Stony Brook, 
	Stony Brook, NY 11794-3800
\thanks{Electronic mail: {\tt Christian.Cardall@sunysb.edu}}}
\date{September 1999}
\maketitle

\begin{abstract}
In the simplistic quantum mechanical picture
of flavor mixing, conditions on the maximum size and 
minimum coherence time
of the source and detector regions for the observation of 
interference---as well as the very viability
of the approach---can only be argued in an ad hoc way
from principles external to the formalism itself.
To examine these conditions in a more fundamental way,
the quantum field theoretical $S$-matrix approach is
employed in this paper, 
without the unrealistic assumption of microscopic stationarity.
The
fully normalized, time-dependent neutrino flavor mixing
event rates presented here automatically reveal the
coherence conditions in
a natural, self-contained, and physically unambiguous way,
while quantitatively describing the transition to their failure.
\end{abstract}


\def\nua{\nu_\alpha}
\def\nub{\nu_\beta}
\def\nuanub{\nu_ \alpha \rightarrow \nu_\beta}
\def\kx{\sum_i k_i}
\def\py{\sum_j p_j}
\def\mat{{\cal M}}
\def\be{\begin{equation}}
\def\ee{\end{equation}}
\def\eqr#1{{Eq.\ (\ref{#1})}}
\def\Xs{{\bf x}_S}
\def\Yd{{\bf y}_D}
\def\eqr#1{{Eq.\ (\ref{#1})}}
\def\eld{\sum_l (-1)^{d_l} E_{\underline {\vp}_l}}
\def\els{\sum_l (-1)^{s_l} E_{\underline {\vk}_l}}
\def\kl{\sum_l (-1)^{s_l} {\underline k_l}}
\def\pl{\sum_l (-1)^{d_l} {\underline p_l}}
\def\pln{\sum_l (-1)^{d_l} p_l}
\def\kln{\sum_l (-1)^{s_l}  k_l}
\def\puln{\sum_l (-1)^{d_l} {\underline p_l}}
\def\kuln{\sum_l (-1)^{s_l} {\underline k_l}}
\def\question#1{{{\marginpar{\tiny \sc #1}}}}
\def\tu#1{{\tilde{\underline #1}}}
\section{Introduction}
\label{sec:intro}
\def\nua{\nu_\alpha}
\def\nub{\nu_\beta}
\def\nuanub{\nu_ \alpha \rightarrow \nu_\beta}
\def\vp{{\bf p}}
\def\vl{{\bf l}}
\def\vk{{\bf k}}
\def\vq{{\bf q}}
\def\vx{{\bf x}}
\def\vy{{\bf y}}
\def\vz{{\bf z}}
\def\vs{{\bf s}}
\def\vr{{\bf r}}
\def\vu{{\bf u}}
\def\vv{{\bf v}}
\def\bp#1{{\underline p_{#1} }}
\def\bk#1{{\underline k_{#1} }}
\def\bvp#1{{\underline \vp_{#1} }}
\def\bvk#1{{\underline \vk_{#1} }}
\def\enu{E_{\vq}}
\def\gmu{\gamma^\mu}
\def\dmu{\partial_\mu}
\def\lhat{\hat{\bf L}}

Several recent works have examined neutrino flavor
mixing by considering the neutrino production/mixing/detection
as a single process in the context of quantum field theory
(QFT) \cite{giunti,grimusstock,camp,kier,giunti2,cohere,ioapilaf,cardall}.
Such a framework clarifies several conceptual difficulties
associated with the familiar quantum mechanical (QM) model of the
flavor mixing process (see Ref. \cite{rich} for a listing of
some of these). One conceptual difficulty
associated with the simplified QM picture is that it postulates
neutrino flavor eigenstates of indefinite mass, 
while in QFT external particles (asymptotic states) are
generally required to be on-shell. Hence the usual methods of calculating
neutrino production rates in QFT would be rates for 
neutrinos of a particular mass, precluding
interference between neutrino states of different mass.
In a QFT description of a neutrino mixing experiment, this problem 
is resolved by considering the neutrinos to be virtual particles. 
After all, it is the measurable, on-shell external charged 
leptons associated with 
the neutrino production and detection processes that operationally
define what is meant by ``neutrino flavor mixing''; the neutrinos
themselves are not directly observed. In the relativistic
limit, the same factors that constitute the ``oscillation amplitude'' 
in the simplified quantum mechanical picture can be identified in the
amplitude for the overall neutrino production/mixing/detection process.

While descriptions of neutrino flavor mixing in QFT have provided 
insight, some shortcomings remain.
As noted in Ref. \cite{cardall}, one problem is that the calculations 
\cite{giunti,grimusstock,camp,kier,giunti2,cohere,ioapilaf}
are not carried out to normalized event rates.
Without normalization, one cannot definitely say that one has identified
an ``oscillation probability.''\footnote{A normalized probability is given
in Ref. \cite{ioapilaf}, but it is not the experimentally relevant one,
which is identified by making a complete connection of the squared
amplitude for the production/mixing/detection process
with the form (neutrino flux)(oscillation probability)
(neutrino cross section)\cite{cardall}.} 
In addition, Refs. 
\cite{giunti,grimusstock,camp,giunti2,cohere} are restricted to particular
neutrino production and detection reactions, while Ref. \cite{kier}
employs idealized two-state systems as source and detector. Since one
would hope to justify the use of the simple QM model in general 
circumstances, such restrictions should not be required. 

Another potential pitfall is a failure to distinguish
between macroscopic stationarity and microscopic 
stationarity.\footnote{This critique applies only to
QFT analyses of mixing experiments, which by their very nature
purport to describe microscopic processes in the source and
detector. Once the feasibility of the use of
flavor eigenstates in a simplified model is established, 
macroscopic stationarity
can be sensibly employed, as in Ref. \cite{stod}.} 
Some previous studies invoke microscopic stationarity,
either implicitly \cite{camp,ioapilaf,cardall}, 
or with explicit reference to bound
states \cite{grimusstock,cohere} in the source and/or detector. 
While sources and detectors
as a whole can sometimes be considered stationary on a macroscopic 
basis, the claim that individual particles in the source and/or detector  
remain unperturbed in coherent states over macroscopic time scales
is dubious.\footnote{The ``macroscopic time scale'' at issue here is the
signal travel time between source and detector. In the case of
astrophysical neutrino sources, this time is macroscopic indeed.}
A good example is the Sun: While macroscopic variables such
as density, pressure, and so on may be stationary, zooming in
to atomic scales one sees a roiling thermodynamic bath of particles
being created, destroyed, and scattered on rapid time scales. 
Clearly there is no hope of appealing to bound states of the nuclei and
electrons which collide to produce neutrinos in the Sun! Even in
detectors which have bound state particles, it is difficult to conceive
of these states as being coherent over macroscopic time
scales. For example, water \v Cerenkov detectors see charged lepton
wave packets with finite energy and time spread (after all, an
``event time'' is recorded whose uncertainty is much smaller than,
for example, the signal travel time between the Sun and Earth). 
While the overall
chain of the processes of detection can be rather complex, one would
expect that at least part of the reason for the finite spread of
these detected charged leptons is the limited coherence time of the 
bound state particles with which the initial detection interaction takes 
place.

In short, if one employs a QFT description of flavor oscillations 
in order to overcome the conceptual difficulties of the simplified QM 
model, one should also pay the price of being 
realistic about the lack of microscopic stationarity in order to 
complete a convincing picture. In this paper, this is accomplished
by treating all of the 
initial and final state external particles as wave
packets which have finite overlap in space and time in the source
and detector. Thus it is similar in spirit to Refs. \cite{giunti,giunti2},
but in pursuit of generality 
the treatment does not specify particular neutrino
production and detection mechanisms or specific functional forms
for the wave packets of the external particles involved in the
production and detection processes. Another difference of the 
present treatment is a greater emphasis on the coordinate space
Green's function, as in Ref. \cite{cardall}, where it was used
to make direct contact with the standard simple 
coordinate space formalism for the
MSW effect. In this work no integration is performed over 
the time coordinate of the detection event,  
since many experiments---including those employing 
water \v Cerenkov detectors---record this time. (An integration
{\em is} performed over the unobserved time coordinate of the
emission event.) 
Finally, the detailed connection of the squared amplitude 
for the microscopic neutrino production/mixing/detection process
to macroscopic event rates will be made. Only this complete
connection---with all factors accounted for---enables one to
define an oscillation probability.

\section{$S$-matrix approach to neutrino mixing processes}

A common application of QFT is the description of particle
collisions in accelerators.
Rates or cross sections associated with these processes can 
be obtained in a heuristic manner directly from the plane wave scattering
$S$-matrix computed from Feynman diagrams,
\begin{equation}
S(\{ p \}  )-1 \equiv (2 \pi)^4 \delta^{4}
\left(\pln\right) i \mat(\{ p \}  ).
\label{smatrix}
\end{equation}
In this expression $\{p\}$ is the set of external particle
momenta $p_l$, $d_l=1$ for incoming and 0 for outgoing particles,
and $\mat$ is the $\delta$ function-free matrix element.
The event rate obtained from Eq. (\ref{smatrix}) is\footnote{The 
conventions for the metric, $\gamma$ matrices,
and normalizations employed here are the same as those of
Ref. \cite{peskin}.}
\begin{equation}
d\Gamma=(2\pi)^4 \delta^{4}\left(\pln\right) V^{1-I}|\mat(\{ p \}  )|^2
\left[\prod_i^I {1\over 2E_{\vp_i}}\right] \left[\prod_{i'}^F 
{d\vp_{i'}\over (2\pi)^3 (2E_{\vp_{i'}})}\right].
\end{equation}
Here $V$ is the three-volume in which the entire process occurs, $I$
and $F$ are the numbers of particles in the initial and final states,
and the components of the on-shell 
four-momenta $p_i$ are $ (E_{\vp_i},\vp_i)$. 
This mnemonic for arriving at event rates 
is possible because the interactions of interest occur 
in a single, small spacetime
volume. It is more convincingly justified, however, by 
a wave packet description (e.g., Ref. \cite{gold}).

One of the reasons one considers a QFT description of
neutrino flavor mixing is that the standard picture of
requiring external particles (asymptotic states)
to be on-shell precludes
the existence of
massive neutrino flavor eigenstates (assuming that each charged
lepton couples to multiple neutrino fields of different masses). 
Accordingly, one
considers the neutrinos as virtual particles in a 
Feynman diagram in which the charged leptons at the source
and detection vertices identify the neutrino flavor.
Neutrino flavor mixing then results from interference of
diagrams whose intermediate neutrinos have different masses.

In this picture it is not possible to compute event rates
directly from the $S$-matrix with the usual mnemonic described
above. This is because 
a neutrino oscillation experiment involves neutrino
production and detection regions which are widely separated in
space. In contrast to the case of accelerator particle collisions,
the interactions of interest do not all occur in a
single volume element. In addition, as argued in Sec. \ref{sec:intro},
in this microscopic picture the production and detection of a single
neutrino will be separated in time as well as space.

In order to describe the spacetime localization 
one must fall back on a wave packet description
of the external particles,
in which the amplitude is a superposition of plane wave amplitudes:
\begin{equation}
{\cal A} = \int \prod_j^{1 + F_D} [dp_j]\
 \phi_{D j}(p_j, \bvp{j})
\prod_i^{I_S+F_S}  [dk_i]\
\phi_{S i}(k_i, \bvk{i})\ 
\left[S(\{ k \}, \{ p \} )-1\right],
\label{amplitude}
\end{equation}
 where (for example)
$[dp_j]=d {\vp}_j/\left[ (2 \pi)^3 \sqrt{2 E_{{\vp}_j}}\right]$, 
$\{ k \}$ 
are
the external momenta connected to the vertex causing neutrino production, and 
$\{ p \}$
are the external momenta connected to 
the vertex associated with neutrino detection.
The quantities 
$\{\underline \vk\}$ and $\{\underline \vp \}$
denote the peak of the wave packets' distribution of
three-momenta. There are $I_S$ incoming and $F_S$ outgoing particles
connected to the production vertex, and 1 incoming and $F_D$ 
outgoing external particles at the
detection vertex. 
The origin of the source  wave packets is taken to be the spacetime
point $x_S$:
\begin{equation}
\phi_{Si}(k_i, \bvk{i}) = a_{Si}(\vk_i- \bvk{i})\,e^{-i(-1)^{d_i} k_i\cdot
x_S},
\end{equation}
and similarly for the detector wave packets with origin $y_D$.
The real function $a(\vk- \underline\vk)$
is peaked about $\underline\vk$.
In order that Eq. (\ref{amplitude}) describe the amplitude for
interaction of one localized external particle of each type,
the wave packet normalization must be, e.g. for the source packets,
\begin{equation}
\int {d{\vk_i}\over (2\pi)^{3}} \left|\phi_{Si}(k_i,\bvk{i})\right|^2~=
\int {d{\vk_i}\over (2\pi)^{3}} \left|a_{Si}(\vk_i-\bvk{i})\right|^2=~1. 
\label{norm1}
\end{equation}
It is convenient to define the transform
\begin{equation}
\psi_{\tiny\bvk{i}}(\vx)=
\int {d{\vk_i}\over (2\pi)^{3}}\, a_{Si}(\vk_i-\bvk{i})\,e^{-i(-1)^{s_i}
\vk\cdot\vx},
\end{equation}
where $s_i$ behaves the same as the $d_l$ of
Eq. (\ref{smatrix}). The normalization
of this function is
\begin{equation}
\int d\vx \left|\psi_{\tiny\bvk{i}}(\vx)\right|^2 = 1,\label{spacenorm}
\end{equation}
which follows from Eq. (\ref{norm1}).

The next step is to transform the momentum-based Eq. (\ref{amplitude})
into a coordinate space expression. The $S$-matrix in Eq. (\ref{amplitude})
can be expressed
\begin{eqnarray}
S(\{ k \}, \{ p \}  )-1& =& \int d^4y\;e^{i \pln \cdot y}
	\int d^4x\;e^{i \kln \cdot x}\,\nonumber \\
& & \times i\int {d^4s\over (2\pi)^4}
	e^{ \mp i s\cdot(y-x)} M_2\, P_L\, G(s)\, P_R\, M_1,
\label{factorit}
\end{eqnarray}
in which 
$s$ is the off-shell neutrino propagator momentum. 
The upper (lower) sign of $\mp$ in the
exponential is for neutrino (antineutrino) mixing.
This arises from choosing $x$ ($y$) to always
correspond to the source (detector). 
That is, for neutrino oscillations of flavor $\alpha$
to flavor $\beta$, the propagator is 
$iG^{\beta\alpha}(y,x)=\langle T\{\nu^\beta(y)\bar\nu^\alpha(x)\}\rangle_0
=i\int d^4s\, (2\pi)^{-4}\,e^{-is\cdot(y-x)}G^{\beta\alpha}(s)$
(with $T\{\}$ and $\langle \rangle_0$ denoting a time-ordered product
and vacuum expectation value respectively),
while for antineutrino oscillations $\alpha\rightarrow \beta$,
the labeling is $iG^{\alpha\beta}(x,y)$. $V-A$ interactions have been
assumed;
$P_L$ and $P_R$ are the left- and right-handed projection operators, with
$M_1$ and $M_2$ column and row vectors in spinor space respectively.
For neutrino mixing, $M_1 = M_1(\{k\})$ and $M_2=M_2(\{p\})$ are respectively
associated with the neutrino production and detection reactions. For 
antineutrino mixing, $M_2 = M_2(\{k\})$ and $M_1 = M_1(\{p\})$ are
respectively associated with the production and detection reactions. 
The partially transformed propagator
$G(s^0,\vy,\vx)$ is defined by
\begin{equation}
\int {d^4s\over (2\pi)^{4}}\,e^{\mp is\cdot(y-x)}G(s)
=\int {ds^0\over 2\pi}\, e^{\mp i s^0(y^0-x^0)}\, G(s^0,\vy,\vx).
\end{equation}
It is assumed that the wave packets $a(\vk-\underline\vk)$ are
sufficiently well-peaked that integrals of the following
form can be evaluated in an approximate manner:
\begin{eqnarray}
\int [dk_i]\, \phi_{Si}(k_i,\bvk{i})\, e^{i(-1)^{s_i}k_i\cdot x}
M(\vk_i)&=& \int {d {\vk}_i\over (2 \pi)^3 \sqrt{2 E_{\tiny{\vk}_i}} }\;
a_{Si}(\vk_i-\bvk{i})\,e^{i(-1)^{s_i}k_i\cdot (x-x_S)}M(\vk_i)\nonumber\\
&\approx& {e^{i(-1)^{s_i}\underline k_i\cdot (x-x_S)}\over
\sqrt{2 E_{\tiny \bvk{i} }} }\;  \psi_{\tiny\bvk{i}}\left((\vx-\vx_S)-
(x^0-x_S^0)\vv_{\tiny\bvk{i}}\right)\, M(\bvk{i}). \label{packet}
\end{eqnarray}
In Eq. (\ref{packet}), $\vv_{\tiny\bvk{i}}$ is the wave packet's
group velocity $(\nabla_{\vk_i} k_i^0)|_{\vk_i={\tiny\bvk{i}}} = 
\bvk{i}/E_{\tiny\bvk{i}}$,
and wave packet spreading has been neglected. Similar expressions
hold for the detector wave packets. With this approximation, 
the amplitude of Eq. (\ref{amplitude}) becomes
\begin{eqnarray}
{\cal A}&=&\left(\prod_i^{I_S+F_S}{1\over \sqrt{2 E_{\tiny\bvk{i}} } }\right)
\left(\prod_j^{1+F_D}{1\over \sqrt{2 E_{\tiny\bvp{j}} } }\right)
\int d^4x\,d^4y\; e^{i\kl\cdot (x-x_S)} e^{i\pl\cdot (y-y_D)}\nonumber\\
& & \times \left[\prod_i^{I_S+F_S}\psi_{\tiny\bvk{i}}\left((\vx-\vx_S)-
(x^0-x_S^0)\vv_{\tiny\bvk{i}}\right) \right]
\left[\prod_j^{1+F_D}\psi_{\tiny\bvp{j}}\left((\vy-\vy_D)-
(y^0-y_D^0)\vv_{\tiny\bvp{j}}\right) \right] \nonumber\\
& & \times i \int {ds^0\over 2\pi}\; e^{\mp i s^0(y^0-x^0)}\, 
{\underline M}_2 P_L \; G(s^0,\vy,\vx)\,P_R {\underline M}_1,
\label{amplitude2}
\end{eqnarray}
in which the bars in ${\underline M}_1$ and ${\underline M}_2$ signify
that these quantities have been evaluated at the peak momenta of
the wave packets.

\section{Approximation of the wave packet overlap}

It is convenient at this stage to adopt an approximation regarding
the overlap of the wave packets that captures the essential physics
while maintaining mathematical simplicity. The initial and final
state wave packets in the source (for example), 
traveling with their various group velocities,
overlap in a limited 
region of space for a limited time. 
To give a specific definition to this spacetime volume of 
the overlap, ${\cal V}_S$, centered on $x_S$,
it is convenient to define
\begin{equation}
E_S(x-x_S)\equiv \left[\prod_i^{I_S+F_S} 
\psi_{\tiny\bvk{i}}\left(\vx_S,x_S^0\right)
  \right]^{-1}
\left[\prod_i^{I_S+F_S}\psi_{\tiny\bvk{i}}\left((\vx-\vx_S)-
(x^0-x_S^0)\vv_{\tiny\bvk{i}}\right)\right],\label{overlap}
\end{equation}
where the notation
\begin{equation}
\psi_{\tiny\bvk{i}}\left(\vx_S,x_S^0\right) =
\left.\psi_{\tiny\bvk{i}}\left((\vx-\vx_S)-
(x^0-x_S^0)\vv_{\tiny\bvk{i}}\right)\right|_{\vx=\vx_S,x^0=x_S^0}
\end{equation}
has been adopted in the first factor.
Then ${\cal V}_S$ is defined by
\begin{eqnarray}
\int d^4x\left[E_S(x-x_S)\right]^2&=&
\int d^4x\exp\left\{2\ln\left[1-{1\over 2}(W_S)_{\mu\nu}(x-x_S)^\mu
(x-x_S)^\nu+\cdots\right]\right\}\nonumber\\
&\approx &  {\pi^2\over\sqrt{{\rm Det}\left[(W_S)_{\mu\nu}\right]}}\equiv
{\cal V}_S, \label{volume}
\end{eqnarray} 
where
\begin{equation}
(W_S)_{\mu\nu}\equiv \left. -{\partial^2 
\over\partial x^\mu \partial x^\nu}E_S(x-x_S)\right|_{x=x_S}.\label{width}
\end{equation}
The timelike (spacelike) components of $(W_S)_{\mu\nu}$ reflect
the spread of energy (momentum) available in the reaction, while
the timelike (spacelike) components of $(W_S^{-1})_{\mu\nu}$ characterize
the extent in time (space) of the wave packet overlap.
Similar considerations apply to the detector region.

It is only necessary here to consider the Green's function
for neutrino propagation through the vacuum. 
Inspection of Ref. \cite{cardall} indicates that 
the vacuum propagator results will be applicable in a relatively
direct
way to the case of 
neutrino propagation through a medium of constant 
density.  
Generalization to the case of a medium of varying density
would be more complicated, however. While 
interference terms for this case 
have been calculated in the context of the simple quantum
mechanical model \cite{beacom}, they are not relevant
to current observations of astrophysical neutrinos. Hence
the effort to study the microscopic origin for the
damping of interference terms already deemed irrelevant  
does not seem to be worthwhile at present.

Focusing on the vacuum case---for which the interference
terms {\em are} of current experimental interest---the 
final factors in Eq. (\ref{amplitude2}) can
be expressed as
\begin{equation}
{\underline M}_2 P_L \; G(s^0,\vy,\vx)\,P_R {\underline M}_1
=\tu{M}_2 \tilde G(s^0,\vy,\vx) \tu{M}{_1}, \label{project}
\end{equation}
where $\tu{M}_2$ and $\tu{M}_1$ are respectively the two-component
subspinors that remain nonzero in $\underline M_2 P_L$ and
$P_R \underline M_1$, and $\tilde G(s^0,\vy,\vx)$ is the 
nonzero $2\times2$
submatrix in $P_L G(s^0,\vy,\vx) P_R$.
Because the overlaps of the wave packets are restricted
to the vicinity of $\vx_S$ and $\vy_D$, and because
$|\vy_D-\vx_S| \gg L_S, L_D$, the leading contribution
from the Green's function is of the form \cite{cardall}
\begin{equation}
\tilde G^{\alpha\beta}(s^0,\vy,\vx)\approx
-\sum_k U_{\alpha k}U_{\beta k}^* \left(s^0 - {s^0\over|s^0|}\,s_k\,
\mbox{\boldmath $\sigma$}\cdot \hat{\bf L}\right){ e^{is_k \hat{\bf L}
\cdot (\vy-\vx)}\over 4\pi |\vy_D-\vx_S|}.\label{green}
\end{equation} 
In this expression, the flavor and mass fields are related by
$\nu_\alpha(x)=\sum_k U_{\alpha k} \psi_k(x)$; 
$s_k = \sqrt{(s^0)^2 - m_k^2}$, in which $m_k$ is the mass associated
with the neutrino field $\psi_k(x)$;
{\boldmath $\sigma$}
is the three-vector of Pauli matrices; and the vector $\hat {\bf L}
=(\vy_D - \vx_S)/|\vy_D-\vx_S|$ points from the source to the detector.
For neutrino oscillations, $\beta$ is the flavor of the charged
lepton associated with the source reaction, and $\alpha$ is the 
flavor of the charged lepton associated with the detection reaction.
For antineutrino oscillations these assignments are reversed.

With these preparations the remaining integrations in Eq. (\ref{amplitude2})
can be performed. Employing Eqs. (\ref{overlap}),
(\ref{project}), and (\ref{green}), and employing a similar approximation
to that employed in Eq. (\ref{volume}) for the $x$ and $y$ integrals,
the amplitude becomes
\begin{eqnarray}
{\cal A}&=&-i\left[\prod_i^{I_S+F_S}
{ \psi_{\tiny\bvk{i}}\left(\vx_S,x_S^0\right)
  \over \sqrt{2 E_{\tiny\bvk{i}} }}\right]
\left[\prod_j^{1+F_D}
{ \psi_{\tiny\bvp{j}}\left(\vy_D,y_D^0\right)
  \over \sqrt{2 E_{\tiny\bvp{j}} }}\right]
{(4 {\cal V}_S)(4{\cal  V}_D)  \over 4\pi |\vy_D-\vx_S|}\nonumber\\
& & \times \sum_k U_{\alpha k}U_{\beta k}^* \int {ds^0\over 2\pi}\;
e^{\mp i s^0(y_D^0-x_S^0) + i s_k |\Yd-\Xs| - D_k(s^0)}
\tu{M}_2\ \left(s^0 - {s^0\over|s^0|}\,s_k\,
\mbox{\boldmath $\sigma$}\cdot \hat{\bf L}\right) \tu{M}{_1},
\label{amplitude3}
\end{eqnarray}
in which the function $\exp[-D_k(s^0)]$,
with
\begin{eqnarray}
D_k(s^0)&=&{1\over 2}(W_S^{-1})_{\mu\nu}
\left(-k_S + \xi_k\right)^\mu \left(-k_S + \xi_k\right)^\nu
\nonumber\\
& &+{1\over 2}(W_D^{-1})_{\mu\nu}
\left(p_D - \xi_k\right)^\mu \left(p_D - \xi_k\right)^\nu \label{delta}
\end{eqnarray} 
enforces energy-momentum conservation to the extent allowed
by the finite overlap in space and time of the external particle
wave packets. In Eq. (\ref{delta}), the notation
\begin{eqnarray}
k_S& \equiv& -\sum_l(-1)^{s_l}\bk{l},\\ 
p_D& \equiv& +\sum_l(-1)^{d_l}\bp{l},\\ 
\xi_k& \equiv& \left(\pm s^0,\hat{\bf L}\sqrt{(s^0)^2-m_k^2} \right)
\end{eqnarray}
has been employed.

If plane wave final states and stationary initial source and/or 
detector particle states had been employed, the finite
energy spread indicated in Eq. (\ref{delta})
would have been replaced by 
an energy delta
function, 
suggesting the idea that the neutrinos are energy eigenstates.
In such a case, one finds
$ s^0 = \pm p_D^0 = \pm k_S^0$ (it will be recalled that
the upper sign is for neutrino emission at the source, and the 
lower sign for antineutrino emission). 
It is easy to see by considering
sample neutrino production and detection processes that $k_S^0$ 
and $p_D^0$ should
be positive quantities. 

While Eq. (\ref{delta}) indicates that a range of $s^0$
contributes to the amplitude, there is a value of 
$s^0$---call it $(\underline s^0)_k$---which makes the
largest contribution to the amplitude. This is the
value of $s^0$ for 
which $D_k(s^0)$ has its minimum value.
The relative degrees to which overall energy and momentum
are conserved compete in determining $(\underline
s^0)_k$. Since the external particles travel at speeds less than the speed
of light, however, the timelike components of the
tensors $(W^{-1}_{S,D})_{\mu\nu}$ will be larger than the spacelike components.
This means that, in analogy with the stationary
situation mentioned above,
$(\underline s^0)_k$
may be taken to be
positive (negative) for neutrino (antineutrino) emission at the source.
This, together with the fact that 
only the region in the vicinity of $(\underline s^0)_k$ will be taken
into account in the approximate evaluation of the integral, means
that the integral over $s^0$ can be replaced by integration over a 
new variable $\lambda$, with $\pm s^0$ replaced by $\lambda$ in the
integrand. This new integral is dominated by the region near 
$\underline \lambda_k \equiv |(\underline s^0)_k|$,
determined by 
\begin{equation}
0 = {dD_k(\lambda)\over d\lambda},\label{neuen}
\end{equation}
in which $D_k(\lambda)$ is given by Eq. (\ref{delta}) with
$\xi_k=\left(\lambda,\hat{\bf L}\sqrt{\lambda^2-m_k^2}\right)$.
Expanding $D_k(\lambda)$ to second order about $\underline 
\lambda_k$, the rest of the argument of the exponential to first order,
the rest of the integrand to zeroth order, performing the $\lambda$
integration, and squaring the amplitude yields the result 
\begin{eqnarray}
|{\cal A}|^2&=&\left[\prod_i^{I_S+F_S}
{\left| \psi_{\tiny\bvk{i}}\left(\vx_S,x_S^0\right)\right|^2
  \over \left(2 E_{\tiny\bvk{i}} \right)}\right]
\left[\prod_j^{1+F_D}
{\left| \psi_{\tiny\bvp{j}}\left(\vy_D,y_D^0\right)\right|^2
  \over \left(2 E_{\tiny\bvp{j}} \right)}\right]
{4({\cal V}_S)^2 ({\cal V}_D)^2 \over \pi^4 
|\vy_D-\vx_S|^2}\left| \sum_k U_{\alpha k}U_{\beta k}^* 
\left(\pi\over \ell_k^2\right)^{1/2}\right.\nonumber\\
& & \times\left.
\exp\left[- i {\underline\lambda_k}(y_D^0-x_S^0) 
+ i s_k(\underline \lambda_k) |\Yd-\Xs| -C_k(\underline\lambda_k,x_S,y_D)
- D_k(\underline\lambda_k)\right]\right.\nonumber\\
& &\times\left.\tu{M}_2\ \left[\underline\lambda_k - s_k(\underline\lambda_k)\,
\mbox{\boldmath $\sigma$}\cdot \hat{\bf L}\right] \tu{M}{_1}\right|^2,
\label{probability}
\end{eqnarray}
where 
\begin{equation}
\ell_k^2 = \left.{1\over 2}{d^2D_k(\lambda)\over d\lambda^2}
\right|_{\lambda=\underline\lambda_k}.
\label{cohere} 
\end{equation}
Study of the explicit expression for $\ell_k^2$ shows that
it is essentially the sum of the squares of the time and length
scales of the wave packet overlaps in the source and detector,
i.e. 
\begin{equation}
\ell_k^2~\sim~(T_S)^2~+~(L_S)^2~+~(T_D)^2~+ (L_D)^2,
\label{cohere2}
\end{equation}
 (in rather obvious
notation). This is particularly transparent in the limit
of relativistic neutrinos.

The factor $\exp[-C_k(\underline\lambda_k,x_S,y_D)]$, with
\begin{equation}
C_k(\underline\lambda_k,x_S,y_D)={1\over 4\ell_k^2}\left[(y_D^0-x_S^0)-
 {1\over v_k}
|\Yd-\Xs|\right]^2,\label{trajectory}
\end{equation} 
suppresses contributions from neutrinos that do not follow a
classical spacetime trajectory between the production event
at $(x_S^0,\vx_S)$ and the detection event at $(y_D^0,\vy_D)$
[the neutrino velocity is given by $v_k = s_k(\underline\lambda_k)/
\underline\lambda_k$].\footnote{The above procedure in which 
a ``neutrino energy'' $\underline \lambda_k$ is determined only from
the minimum of $D_k(\lambda)$ implicitly assumes that the
phase $-\lambda (y_D^0-x_S^0) + s_k(\lambda)|\vy_D-\vx_S|$ 
in Eq. (\ref{amplitude3}) is essentially
stationary over the range of $\lambda$ for which $\exp[-D_k(\lambda)]$
is appreciably nonzero. It is easy to see that the quantity
$\exp[-C_k(\underline\lambda_k,x_S,y_D)]$ enforces this very 
condition, so that the procedure is self-consistent.}
  
\section{Macroscopic event rate}

To make contact with experiments it is necessary to magnify
the probability of Eq. (\ref{probability}) up to macroscopic
scales. For this purpose, the normalization in Eq. (\ref{spacenorm}) 
suggests that (for example) 
$\left| \psi_{\tiny\bvk{i}}\left(\vx_S,x_S^0\right)\right|^2$
be interpreted as the (per particle) volume density of particles
with momentum $\bvk{i}$ at position $\vx_S$ and time $x_S^0$, where
the last two quantities are now thought of as macroscopic
spacetime variables. Employing the usual statistical methods
for free particles, these particle
densities are taken to be $[d\bvk{i}/(2\pi)^3]\,f(\bvk{i},\vx_S,x_S^0)$ 
for initial state
particles (where $f$ is the phase space density)
and $[d\bvk{i}/(2\pi)^3]$ for final state particles.
In connection with the (now macroscopic) variables $\vx_S$, $x_S^0$,
$\vy_D$, and $y_D^0$, one factor of $({\cal V}_S {\cal V}_D )$ is 
interpreted as $d\vx_S\,dx_S^0\,d\vy_D\,dy_D^0$. At the macroscopic 
level, a sum over external particle spins is performed; 
the average over initial spins is accounted for by leaving the
spin degeneracy out of the phase space distribution functions $f$.
The expected number of events detected from neutrino interactions with the
$[d\bvp{}/(2\pi)^3] f(\bvp{},\vy_D,y_D^0)$ particles of momentum
$\bvp{}$ in detector
volume $d\vy_D$ during time $dy_D^0$ resulting in final state 
detector particles
of momentum $\{\bvp{j'}\}$, arising from neutrinos produced from
the interaction of the set of $\{[d\bvk{i}/(2\pi)^3] f(\bvk{i},\vx_S,x_S^0)\}$ 
detector particles with momenta $\{\bvk{i}\}$
in source volume $d\vx_S$ during time $dx_S^0$
resulting in final state source particles of momentum $\{\bvk{i'}\}$, is
\begin{eqnarray}
dN&=&d\vx_S\;dx_S^0\;d\vy_D\;dy_D^0\;
d{ {\bf K}(x_S)}\; d{ {\bf K}}'\; d{ {\bf P}(y_D)}\; d{ {\bf P}}'\;
{4{\cal V}_S {\cal V}_D \over \pi^4 |\vy_D-\vx_S|^2}
 \sum_{\rm spins}
\left| \sum_k U_{\alpha k}U_{\beta k}^* 
\left(\pi\over \ell_k^2\right)^{1/2}\right.\nonumber\\
& &\times\left.
\exp \left[- i {\underline\lambda_k}(y_D^0-x_S^0) 
+ i s_k(\underline \lambda_k) |\Yd-\Xs| -C_k(\underline\lambda_k,x_S,y_D)
- D_k(\underline\lambda_k)\right]\right.\nonumber\\
& &\times\left.\tu{M}_2\ \left[\underline\lambda_k - s_k(\underline\lambda_k)
\mbox{\boldmath $\sigma$}\cdot \hat{\bf L}\right] \tu{M}{_1}\right|^2,
\label{events}
\end{eqnarray}
where the notation
\begin{eqnarray}
d{ {\bf K}(x_S)}&=&\prod_i^{I_S}
 {d\bvk{i}\over (2\pi)^3 \left(2 E_{\tiny\bvk{i}} \right)}\;
f\left(\bvk{i},\vx_S,x_S^0\right),\\
d{ {\bf K}}'&=&\prod_{i'}^{F_S}
 {d\bvk{i'}\over (2\pi)^3 \left(2 E_{\tiny\bvk{i'}} \right)},\\
d{ {\bf P}(y_D)}&=&
 {d\bvp{}\over (2\pi)^3 \left(2 E_{\tiny\bvp{}} \right)}\;
f\left(\bvp{},\vy_D,y_D^0\right),\\
d{ {\bf P}}'&=&\prod_{j'}^{F_D}
 {d\bvp{j'}\over (2\pi)^3 \left(2 E_{\tiny\bvp{j'}} \right)}
\end{eqnarray}
has been introduced for the phase space factors.

While virtually all neutrino experiments record data
that sums over contributions from all initial momenta
in the source and detector, all source final momenta, and all
source emission times,    
some experiments---such as those with water \v Cerenkov 
detectors---record the detector event time and (at least some) 
detector final state particle momenta. Hence it would not be
correct to integrate over these last quantities. 
Also, dividing by $dy_D^0$ gives an expected
event rate as a function of detector time $y_D^0$.

In the integration over all source times $x_S^0$, the
formalism automatically ``knows'' that neutrinos emitted
at macroscopically different source times are not allowed to
interfere coherently.
Each term in the squared sum
of the form $|\sum_k h(k)|^2=\sum_k\sum_{k'} h(k) h^*(k')$ in
Eq. (\ref{events}) has a factor $\exp[-{\cal T}_{kk'}]$, where
\begin{equation}
{\cal T}_{kk'}=-i(\underline\lambda_k-\underline\lambda_{k'})
(y_D^0-x_S^0) - C_k(\underline\lambda_k,x_S,y_D) -
C_{k'}(\underline\lambda_{k'},x_S,y_D). \label{timearg}
\end{equation}
Terms with $k\ne k'$ represent quantum interference terms.
For a given time of detection $y_D^0$, 
$\exp[-C_k(\underline\lambda_k,x_S,y_D)]$ and
$\exp[-C_{k'}(\underline\lambda_{k'},x_S,y_D)]$ tend to 
pick out different emission times for $k\ne k'$. If the
difference in emission times is greater than the width
of $\exp[-C]$, the interference is suppressed.

The gradual loss of coherence can be expressed quantitatively.  
The leading contribution to terms with $k\ne k'$ comes 
from the interval near 
the average emission time
\begin{equation}
(\underline x_S^0)_{kk'} = y_D^0 -{1\over \underline v_{kk'}} |\vy_D-\vx_S|,
\label{xave}
\end{equation}
where $C_k(\underline\lambda_k,x_S,y_D)+
C_{k'}(\underline\lambda_{k'},x_S,y_D)$ has a minimum. The
``average velocity'' $\underline v_{kk'}$ is given by
\begin{equation}
\underline v_{kk'}={v_k v_{k'} \left( \ell_k^2  +\ell_{k'}^2\right)
 \over \left( \ell_k^2 v_k +\ell_{k'}^2 v_{k'}   
\right)}.
\end{equation} 
The portion of the argument of the exponential that depends on
$x_S^0$---that is, ${\cal T}_{kk'}$ of Eq. (\ref{timearg})---can be expressed
\begin{eqnarray}
{\cal T}_{kk'}&=&{-i(\underline\lambda_k-\underline\lambda_{k'})\over
\underline v_{kk'}}|\vy_D-\vx_S| - {(v_k-v_{k'})^2\over
4v_k^2 v_{k'}^2 \left( \ell_k^2 v_k +\ell_{k'}^2 v_{k'}   \right)}
|\vy_D-\vx_S|^2
\nonumber\\
& & \nonumber\\
& &-{(\underline\lambda_k-\underline\lambda_{k'})^2 \ell_k^2\ell_{k'}^2 \over
\left( \ell_k^2  +\ell_{k'}^2    \right)} - 
{\left( \ell_k^2  +\ell_{k'}^2    \right)\over 4 \ell_k^2\ell_{k'}^2}
\left[x_S^0 - (\underline x_S^0)_{kk'} - {2 i \ell_k^2\ell_{k'}^2
(\underline\lambda_k-\underline\lambda_{k'})\over
\left( \ell_k^2  +\ell_{k'}^2    \right)}\right]^2.\label{timearg2}
\end{eqnarray}
The second term yields an exponential fall off with
$|\vy_D-\vx_S|^2$
in interference between
neutrinos with different masses (and hence different 
velocities).\footnote{Unlike the case of Eq. (\ref{amplitude3}) 
in which there was a self-consistent way to assume a stationary
phase, here it is necessary to include the phase in completing
the square for $x_S^0$. It is this which gives rise to the
second term in Eq. (\ref{timearg2}), which causes a loss
of coherence with increasing $|\vy_D-\vx_S|^2$. 
It can be shown that the
exponential fall off in $|\vy_D-\vx_S|^2$ also ensures that the
phase $-\lambda (y_D^0-x_S^0) + s_k(\lambda)|\vy_D-\vx_S|$ 
in Eq. (\ref{amplitude3}) remains 
stationary over the range of $\lambda$ for which $\exp[-D_k(\lambda)]$
is appreciably nonzero,  for $x_S^0$ determined by the average
velocity $\underline v_{kk'}$ [see Eq. (\ref{xave})]. It will be
recalled that the stationarity of this phase ensures
that the minimum of $D_k(\lambda)$ dominates the integral
(i.e., the ``neutrino energy'' becomes a meaningful concept).}
Setting Eqs. (\ref{timearg}) and (\ref{timearg2}) into Eq. (\ref{events}),
integrating over $x_S^0$ and the unobserved external momenta,
and dividing by $dy_D^0$ gives the expected event rate in the detector
at time $y_D^0$.

\section{Relativistic Limit}

Here the event rate in the limit of
relativistic neutrinos will be exhibited.
The zeroth order neutrino energy $\underline \lambda$ and coherence width 
$\ell$ are given by
Eqs. (\ref{neuen}) and (\ref{cohere}) respectively, with $m_k=0$.
To first order,
\begin{eqnarray}
\underline\lambda_k& =& \underline\lambda +\delta\underline\lambda_k,
\nonumber\\
s_k(\underline\lambda_k)&=&\underline\lambda +
\delta\underline\lambda_k
-{m_k^2\over 2\underline\lambda},\nonumber\\
v_k& =& 1 - {m_k^2\over 2\underline\lambda^2},\nonumber\\
\ell_k^2& =& \ell^2 +\delta\ell_k^2,\nonumber\\ 
\underline v_{kk'}&=& 1-{(m_k^2+m_{k'}^2)\over 4\underline\lambda^2}, 
\label{relapprox}
\end{eqnarray}
where the explicit forms for $\delta\underline\lambda_k$ and 
$\delta\ell_k^2$ are determined
by Eqs. (\ref{neuen}) and (\ref{cohere}) respectively.
It will be assumed that $m_k^2$ can be neglected everywhere
except when appearing with the macroscopic distance
$|\vy_D-\vx_S|$ in the argument of
the exponential, which magnifies its impact. 
This means that the third term in Eq. (\ref{timearg2})
can be neglected, and that 
\begin{equation}
D_k(\underline\lambda_k)\approx D(\underline\lambda),
\end{equation}
where $D(\underline\lambda)$ is given by Eq. (\ref{delta}) with
$\xi_k$ replaced by $\xi\equiv \left(\underline\lambda,
\underline\lambda\hat{\bf L}\right)$.
It also means that 
\begin{eqnarray}
\tu{M}_2\ \left[\underline\lambda_k - s_k(\underline\lambda_k)
\mbox{\boldmath $\sigma$}\cdot \hat{\bf L}\right] \tu{M}{_1}
&\approx& \tu{M}_2\ \left[\underline\lambda - \underline\lambda
\mbox{\boldmath $\sigma$}\cdot \hat{\bf L}\right] \tu{M}{_1}\nonumber\\
&=& \left[{\cal M}_S\left(\{\bvk{}\},\underline\lambda\right)\right] 
\left[{\cal M}_D\left(\{\bvp{}\},\underline\lambda\right)\right],
\end{eqnarray}
where ${\cal M}_S$ and 
${\cal M}_D$ are the 
$\delta$ function-free matrix elements that would appear in
the plane wave $S$-matrices [see Eq. (\ref{smatrix})] 
describing the source and detector reactions with a massless neutrino
of momentum $\underline \lambda\hat{\bf L}$, with the other 
particles having momenta $\{\bvk{}\}$ (source) and $\{\bvp{}\}$
(detector) \cite{cardall}. 
The event rate at detector time $y_D^0$ is
\begin{eqnarray}
d\Gamma(y_D^0)&=&d{ {\bf P}}' \int d\vx_S \int d\vy_D\;
\int \left.d{ {\bf K}(x_S)}\right|_{x_S^0=\underline x_S^0} 
\int d{ {\bf K}}' \int d{ {\bf P}(y_D)}  
{8{\cal V}_S{\cal V}_D\over \pi^3 |\vy_D-\vx_S|^2}\nonumber\\
& &\times \left[\sum_{\rm spins}\left|{\cal M}_S\left(\{\bvk{}\},
\underline\lambda\right)\right|^2\right]
 \left[\sum_{\rm spins}\left|{\cal M}_D\left(\{\bvp{}\},\underline\lambda
\right)\right|^2\right]
\left(\pi\over 2\ell^2\right)^{1/2} e^{-2 D(\underline\lambda)} 
\nonumber\\
& &\times
 \sum_{k,k'} U_{\alpha k}U_{\beta k}^* U_{\alpha k'}^* U_{\beta k'} 
\exp \left[- i {(m_k^2 - m_{k'}^2)|\vy_D-\vx_S|\over 2\underline\lambda}
-{(m_k^2 - m_{k'}^2)^2|\vy_D-\vx_S|^2\over 32 \underline\lambda^4 \ell^2}
\right].
\label{events2}
\end{eqnarray}
It can be shown that 
\begin{equation}
\left(\pi\over 2\ell^2\right)^{1/2} e^{-2 D(\underline\lambda)}
=\int d\enu\; e^{-2 D(\enu)},
\end{equation}
and consistency with the earlier approximate evaluation of the
$s^0$ (or $\lambda$) integration means that $\left|{\cal M}_S\left(\{\bvk{}\},
\underline\lambda\right)\right|^2$,
$\left|{\cal M}_D\left(\{\bvp{}\},\underline\lambda
\right)\right|^2$, and the factors summed over $k,k'$ 
can be taken inside this integral as
functions of $\enu$ rather than $\underline\lambda$.
In addition, if the phase space densities change little
with energy variations 
and momentum variations of order $\ell^{-1}$,
then the leading contribution to Eq. (\ref{events2}) is the
same as if the replacement
\begin{eqnarray}
e^{-2 D(\enu)}& \rightarrow &{\pi^4\over 
\sqrt{{\rm Det}\left[(W_S^{-1})_{\mu\nu}\right]}
\sqrt{{\rm Det}\left[(W_D^{-1})_{\mu\nu}\right]} }\,
\delta^4\left(-k_S + q\right)
\delta^4\left(p_D - q\right)\nonumber\\
&=&{\pi^8\over{\cal V}_S {\cal V}_D}\,
\delta^4\left(-k_S + q\right)
\delta^4\left(p_D - q\right)\nonumber\\
\end{eqnarray}
had been made, where $q = (\enu,\enu\,\hat{\bf L})$.
Hence the leading contribution to the macroscopic event rate
in the detector at time $y_D^0$ can be expressed
\begin{eqnarray}
d\Gamma(y_D^0)&=&
\int d\vx_S \int d\vy_D\int\left[\prod_i^{I_S}
 {d\bvk{i}\over (2\pi)^3}\right]\;
\left[\left. 
f\left(\bvk{i},\vx_S,x_S^0\right)\right|_{x_S^0=y_D^0-|\vy_D-\vx_S|}
\right]\nonumber\\
& &\times
\int {d\bvp{}\over (2\pi)^3 }\;
f\left(\bvp{},\vy_D,y_D^0\right)
\; d\Gamma\left(\{\bvk{}\},\{\bvp{}\},\vx_S,\vy_D\right),
\label{events3}
\end{eqnarray}
where the single particle event rate  is 
\begin{eqnarray}
d\Gamma\left(\{\bvk{}\},
\{\bvp{}\},\vx_S,\vy_D\right)&=&\int dE_{\vq} 
 \left[d\Gamma\left(\{\bvk{}\},\enu\right) \over
	|\vy_D-\vx_S|^2\, d\Omega_{\vq} \, dE_{\vq} \right]\nonumber\\ 
& &\times	\left[P_{\rm mix}\left(\enu,\vx_S,\vy_D\right)\right]
	\left[d\sigma\left(\{\bvp{}\},\enu\right)\right]. \label{rate}
\end{eqnarray}
In Eq. (\ref{rate}), 
\begin{eqnarray}
dE_{\vq} 
 \left[d\Gamma\left(\{\bvk{}\},\enu\right) \over
	|\vy_D-\vx_S|^2\, d\Omega_{\vq} \, dE_{\vq} \right]&=&
{1\over |\vy_D-\vx_S|^2}{\enu^2\, d\enu\over(2\pi)^3 (2\enu)}
\left[\prod_i^{I_S}{1\over\left(2 E_{\tiny\bvk{i}} \right)}\right]
\left[\prod_{i'}^{F_S}\int
 {d\bvk{i'}\over (2\pi)^3 \left(2 E_{\tiny\bvk{i'}} \right)}\right]
\nonumber\\
& &\times\sum_{\rm spins}\left|{\cal M}_S\left(\{\bvk{}\},
\enu\right)\right|^2\;(2\pi)^4
\delta^4\left(-k_S + q\right)
\end{eqnarray}
 is the flux of neutrinos
of energy $\enu$ at position $\vy_D$ due to an interaction
at $\vx_S$, as would be computed with standard plane wave methods; 
\begin{eqnarray}
d\sigma\left(\{\bvp{}\},\enu\right)&=&
{1\over\left(2\enu\right)\left(2 E_{\tiny\bvp{}} \right)}
\left[\prod_{j'}^{F_D}
 {d\bvp{j'}\over (2\pi)^3 \left(2 E_{\tiny\bvp{j'}} \right)}\right]
\nonumber\\
& &\times 
\sum_{\rm spins}\left|{\cal M}_D\left(\{\bvp{}\},
\enu\right)\right|^2\;(2\pi)^4
\delta^4\left(p_D - q\right)
\end{eqnarray}
is the cross section for a massless neutrino interaction
in the detector (assuming nonrelativistic initial state
detector particle momentum
$\bvp{}$, so that the M{\o}ller velocity is equal to 1), and
\begin{eqnarray}
P_{\rm mix}\left(\enu,\vx_S,\vy_D\right)& =& 
\sum_{k,k'} U_{\alpha k}U_{\beta k}^* U_{\alpha k'}^* U_{\beta k'} 
\nonumber\\ & &\times
\exp \left[- i {(m_k^2 - m_{k'}^2)|\vy_D-\vx_S|\over 2\enu}
-{(m_k^2 - m_{k'}^2)^2|\vy_D-\vx_S|^2\over 32 \enu^4 \ell^2}
\right]\label{oscprob}
\end{eqnarray}
is the flavor mixing (or ``oscillation'') probability.

Except for two differences, 
Equations (\ref{events3})-(\ref{oscprob}) are just what one
would write down for a macroscopic event rate using the
naive QM model of the neutrino flavor mixing process, 
together with elementary
considerations for the production flux and detection cross 
sections. The first difference is one that 
could also have been put in by
hand, namely, the causal time delay between emission and
detection.\footnote{In setting $x_S^0=y_D^0-|\vy_D-\vx_S|$ 
in Eq. (\ref{events3}), it has
been assumed that the phase space densities $f$ vary on time
scales slower than $8\times 10^{-19}~{\rm s}\  
[(m_k^2~+~ m_{k'}^2)~/~{\rm eV}^2]({\rm MeV}^2~/~\enu^2)
(|\vy_D-\vx_S|~/~{\rm km})$, for all $k$ and $k'$, in order that
the phase space densities could be taken out of the sum.} 
The second difference is the damping of coherence
at very large distances, discussed earlier in this section.

\section{Discussion}
\label{sec:disc}

The calculations presented here concern the description of
neutrino flavor mixing as a quantum field theoretic process, with the
neutrinos as virtual particles connecting the 
``on shell'' external particles involved in the neutrino production and 
detection reactions. This framework provides a more realistic
description of and deeper physical insight into the flavor mixing process
than the naive quantum mechanical model.
Development beyond previous works has been sought in this study
by considering general
production and detection processes and wave packet functional forms, 
avoiding the unrealistic assumption
of microscopic stationarity, leaving the time of the detection
event (but not the emission event) an observable quantity, 
and making a complete connection 
to fully normalized event rates. The final results are given in 
Eqs. (\ref{events3})-(\ref{oscprob}). 
Note that Eq. (\ref{oscprob}) has a damping
factor for interference terms in addition to the usual oscillatory 
factor; this should formally be considered part of the ``oscillation
probability,'' which (without ambiguity, in the present 
fully normalized treatment) includes everything outside of the 
neutrino emission flux and detection cross section.

The free external particle wave packet picture is convenient for
a number of reasons. Unlike descriptions involving bound states in
the source, it is a
suitable description for astrophysical neutrino sources such as the 
Sun (modulo the Coulomb repulsion of reacting nuclei, which can be
suitably included) or supernovae. 
The lack of stationarity arises naturally due to the finite duration
of the wave packets' overlap. 
The matter of normalization is simple since free particle states are
employed. Furthermore, the dynamics are already built into the $S$-matrix,
making the description of neutrino oscillations a matter of working
out the kinematics.

This $S$-matrix framework could be 
generalized without much difficulty to include bound states for some of 
the source and detector particles. 
(The remaining  particles would still be considered free particle wave
packets. In this framework,
the coherence times of the source and
detector---which ultimately result from complicated microscopic
many-body
physics not considered here---can be considered as input parameters
which ultimately manifest themselves in the finite free particle
wave packet sizes.)
The analogue of the $S$-matrix would be the amplitude
for particular plane wave states to interact with
particular bound states. A superposition
of such amplitudes over several plane wave momenta---in order to
create time-dependent wave packets for the external free particles---would 
constitute the
amplitude for the overall neutrino production/propagation/detection
process, with the neutrino production and detection localized 
in space {\em and} time. 
 In
going to the macroscopic rates, the square of bound state wave functions
would be replaced by a sum over the phase space distribution of
the relevant bound state quantum numbers.

While not new to this study, 
three basic insights into the neutrino flavor mixing process
are listed
here for completeness.
First, {\em (1) an ``oscillation probability'' independent of the
details of production and detection can only be defined in
the relativstic limit.} This limit allows the neutrinos to become
effectively on shell (i.e. massless) as far as production and
detection are concerned. Assuming chiral interactions, the
relativistic limit also causes
only one neutrino spin to contribute, so that the overall squared
amplitude $|{\cal M}|^2$ can factorize into separate production
and detection squared amplitudes $|{\cal M}_S|^2$ and $|{\cal M}_D|^2$. 
This process has been shown in detail in this paper, culminating
in Eqs. (\ref{events3})-(\ref{oscprob}).

The second insight is a condition on {\em (2) the maximum size
of the external particle coordinate space wave function overlap 
in the source ($L_S$) and detector ($L_D$) that allows neutrino
states of different mass to interfere coherently:}
\begin{equation}
L_S,\ L_D \lesssim 0.2\ {\rm m}\  
\left(E_\nu\over {\rm MeV}\right)\left({\rm eV}^2
\over|m_k^2 - m_{k'}^2|\right)
 \equiv L_{\rm osc}/(4\pi).
\end{equation}
When expressed in terms of the ``oscillation length'' this
condition is intuitively obvious.
Its necessity can be seen mathematically
in Eqs. (\ref{delta}) and (\ref{probability}).
From these equations one can see that
the ``neutrino energy'' is determined by a compromise between the
degrees to which energy and momentum are conserved in the source
and detector. However, the tendency towards
energy conservation generally has a greater impact,
i.e. the timelike components $(W_S^{-1})_{00}\sim (T_S)^2$ and 
$(W_D^{-1})_{00}\sim (T_D)^2$ (where $T_S,\ T_D$ are the time scales
of the wave packet overlaps in the source and detector) 
are larger than the spacelike $(W_S^{-1})_{ii}\sim (L_S)^2$  
and $(W_D^{-1})_{ii}\sim (L_D)^2$ 
since the external particles travel slower than the speed of light.
To the extent that the energy is more well determined, there must
be a greater spread in momentum in order for interference to
occur, which is why the condition expressed
above is couched in terms of the spatial (as opposed to temporal) spread
of the wave packet overlap.

The finite duration of the production and detection processes
leads to the third insight. Contributions to the amplitude
from neutrinos that deviate from a classical spacetime trajectory
are exponentially suppressed [see Eqs. (\ref{probability}) and 
(\ref{trajectory})]. This leads to
{\em (3) an upper limit on the number of observable oscillations 
in space:} 
\begin{equation}
N_{\rm osc}={|\vy_D-\vx_S|\over L_{\rm osc}}\lesssim
\pi^{-1}E_\nu \ell \sim
{E_\nu\over \Delta E_\nu}
\end{equation}
[see Eq. (\ref{oscprob})].
Here the ``coherence width'' $\ell\sim\sqrt{(T_S)^2+(T_D)^2+(L_S)^2
+(L_D)^2}$ [see Eqs. (\ref{cohere}) and 
(\ref{cohere2})], and
the detector resolution $\Delta E_\nu = \Delta p_D^0$ has been taken as a
crude estimate of this quantity. Thus many
oscillations in space should be visible before decoherence sets in
as the spatial {\em and} temporal\footnote{Note that if $T_S$ or 
$T_D\rightarrow \infty$, coherence is restored for infinite
propagation distances. This dependence on the coherence time
in addition to the spatial extent of the detector was noted
explicitly in Refs. \cite{kier} and \cite{giunti2}, though it
was implicitly present in Ref. 
\cite{giunti} as well.} resolution of the detector 
begins to distinguish
the separating neutrino mass eigenstates. 
 
It is true that the three insights above can, to some extent,
be achieved without the elaborate machinery presented here.
As far as insight (1) goes, common knowledge that weak interactions
are $V-A$ makes the irrelevance of the neutrino spin degree of freedom
in the relativistic limit somewhat obvious. The relativistic
limit also makes the notion of ``real'' flavor eigenstates 
reasonable (zero mass is 
on-shell). One can then 
adopt the simplified quantum mechanical picture. 
In connection with this simplified quantum mechanical picture, 
condition (2) can be argued from the uncertainty principle,
$\Delta x\Delta p \gtrsim 1$, where $\Delta x$ corresponds
to the oscillation length $L_{\rm osc}$ and $\Delta p$ corresponds to
the inverse source/detector sizes $L_{S,D}^{-1}$ \cite{kays}. 
Condition (3) follows from noting
that a real source (detector) will have some finite 
linewidth (resolution). In order that interference terms
not wash out when binned over this energy range $\Delta E$, 
it is necessary (e.g., \cite{stod})
that the variation in the oscillation phase,
$\Delta[(m_k^2-m_{k'}^2)|\vy_D-\vx_S|/E] \sim (|\vy_D-\vy_S|/L_{\rm osc})
(\Delta E/E)$, be smaller than $2\pi$, which is essentially 
condition (3).

Even if these arguments can be made in some fashion 
in connection with
the simplistic formalism, the whole picture lends itself 
to conceptual difficulties at some level \cite{rich}.
Furthermore, the conditions (1)-(3) must be invoked from
principles outside the formalism itself. In contrast,
the QFT description of the neutrino production/propagation/detection
presented here exhibits all of these conditions in a natural,
self-contained, and physically unambiguous manner, 
and describes the 
transition to the failure of these conditions in a quantitative way.

\acknowledgements{ It is a pleasure to thank Daniel J. H. Chung
for instructive discussions and patient collaboration. 
Discussions with John Beacom and Georg Raffelt  
 are also gratefully acknowledged.
This work
was completed at the GAAC sponsored meeting held at the
Max Plank Institute for Physics and Max Plank Institute for
Astrophysics. CYC
is supported by DOE grant FG02-87ER40317.}


\end{document}